\documentclass[12pt]{iopart}

\usepackage{iopams}

\expandafter\let\csname equation*\endcsname\relax

\expandafter\let\csname endequation*\endcsname\relax
\usepackage{comment}
\usepackage{amsmath}
\usepackage{cite} 
\usepackage{tikz}
\usepackage{graphicx}
\usepackage{soul}
\usetikzlibrary{matrix,shapes,arrows,positioning,chains}

\newtheorem{theorem}{Theorem}[section]

\begin{document}

\title{Soliton-like behaviour in non-integrable systems}
\author{Raghavendra Nimiwal, Urbashi Satpathi, Vishal Vasan, Manas Kulkarni}

\address{International centre for theoretical sciences, Tata Institute of Fundamental Research, Bangalore - 560089, India}

\ead{raghavendra.nimiwal@icts.res.in, urbashi.satpathi@icts.res.in, vishal.vasan@icts.res.in, manas.kulkarni@icts.res.in }
\vspace{10pt}
\begin{abstract}
We present a general scheme for constructing robust excitations (soliton-like) in non-integrable multicomponent systems. By robust, we mean localised excitations that propagate with almost constant velocity and which interact cleanly with little to no radiation. We achieve this via a reduction of these complex systems to more familiar effective chiral field-theories using perturbation techniques and the Fredholm alternative. As a specific platform, we consider the generalised multicomponent Nonlinear Schr\"{o}dinger Equations (MNLS) with arbitrary interaction coefficients. This non-integrable system reduces to uncoupled Korteweg-de Vries (KdV) equations, one for each sound speed of the system. This reduction then enables us to exploit the multi-soliton solutions of the KdV equation which in turn leads to the construction of soliton-like profiles for the original non-integrable system. We demonstrate that this powerful technique leads to the coherent evolution of excitations with minimal radiative loss in arbitrary non-integrable systems. These constructed coherent objects for non-integrable systems bear remarkably close resemblance to true solitons of integrable models. Although we use the ubiquitous MNLS system as a platform, our findings are a major step forward towards constructing excitations in generic continuum non-integrable systems. 
\end{abstract}

%
%
%
%
%

\section{Introduction and summary of results}


Nonlinear dynamics has been a subject of great interest both from a theoretical \cite{jackson1989perspectives, morsch2006dynamics} and an experimental~\cite{dutton2001observation, khaykovich2002formation,  joseph2007measurement,  joseph2011observation} perspective. Such far-from-equilibrium dynamics can be divided into two broad classes: (i) integrable~\cite{das1989integrable,babelon2003introduction} and (ii) non-integrable (generic) \cite{thompson2002nonlinear, tabor1989chaos} models.  If one restricts oneself to conservative systems, then the two hallmarks of far-from-equilibrium dynamics are nonlinearity and dispersion~\cite{kulkarni2012hydrodynamics,tao2006nonlinear}. Nonlinearity is dominant when the dynamics is far from its equilibrium state. For instance, if one starts with initial conditions which are a significant perturbation on top of a background state, we expect nonlinearity to play a role. Such dynamics cannot be merely described by linearisation which will yield a wave equation. Dispersive terms on the other hand are dominant when the profiles have large higher-order spatial derivatives. 

Therefore the far-from-equilibrium dynamics of an initial-value problem can be viewed as an intricate interplay between nonlinearity and dispersion. Nonlinear terms tend to deform the profiles and dispersive terms tend to generate oscillations. In integrable models, there are a class of very special initial conditions (solitons~\cite{drazin1989solitons,ablowitz1981solitons}) where there is a perfect balance between nonlinearity and dispersion.  For solitons, the balance between the two is so remarkably perfect that the initial condition moves robustly with a certain soliton velocity without changing its shape. The situation with non-integrable models is entirely different. In this case, there is an interesting interplay between nonlinear and dispersive terms such that initial profiles typically get deformed and break into oscillations during the course of their evolution.

%

A natural question one can ask is whether, even in non-integrable models, there is a class of special initial conditions which result in solutions that bear close resemblance to solitons of integrable models. By `close resemblance' we mean whether there exist localised excitations that (a) propagate with a constant speed, (b) interact cleanly with one another, namely with no radiation generated. Crudely thinking, one may suppose the answer is a tentative `maybe' for the former requirement and `probably not' for the latter. This is because, one expects non-integrable models (which are rather generic) to lack the necessary mathematical structure to support our two requirements for soliton-like behaviour. In this paper, we show  via perturbation techniques and the Fredholm alternative that one can readily obtain analytic expressions for initial conditions that result in localised constant speed excitations. Furthermore, these objects interact with remarkably minimal radiative loss. Thus we answer the question of whether special initial conditions exist to generate soliton-like behaviour in the affirmative. Of course, one may look for constant-speed excitations by employing a travelling wave Ansatz and some numerical methods to solve nonlinear ordinary-differential equations in combination with a numerical continuation package \cite{doedel1981auto,allgower2012numerical}. However, unlike our prescription, such methods are unlikely to yield explicit expressions that are interpretable and transparently relate the physical parameters to the initial condition.

We wish to address the problem of identifying initial conditions that give rise to soliton-like behaviour for a sufficiently broad class of non-integrable models. Thus we introduce a suitable platform which comprises of a family of non-integrable models. The choice of our platform is based on certain criterion -  (i) A sufficiently generic family of non-integrable models that are ubiquitous in physics and mathematics (ii) Potential to realise the model experimentally and (iii) Potential to design / fabricate initial conditions and make subsequent measurements of time evolution. A platform that fits well with the aforementioned criterion is the multicomponent Nonlinear Schr\"{o}dinger Equations (MNLS, Eq.~\ref{mnls}) also referred to as  multicomponent Gross-Pitaevskii Equations. 

MNLS in various forms and limits has been widely studied in both physics and mathematics. It contains in it special integrable cases and possesses fascinating generalisations including discrete integrable \cite{ablowitz2004discrete, ablowitz1976nonlinear}, discrete non-integrable \cite{kevrekidis2009discrete},  non-local ~\cite{ablowitz2013integrable, ablowitz2016inverse, ablowitz2017integrable} and even disordered cases \cite{scharf1992nonlinear} to name a few.  Various limits of MNLS have been realized in cold atomic physics~\cite{dalfovo1999theory,everitt2015observation,inouye1998observation,xi2016droplet,abdullaev2001stability,zakeri2018solitons,ramesh2010phase}, nonlinear optics~\cite{kivshar2003optical,agrawal2000nonlinear,pushkarov1996bright,skarka1997spatiotemporal,boudebs2003experimental}, plasma physics \cite{lee2007resonant} and molecular systems \cite{davydov1985solitons} to name a few. Moreover, the interaction strengths are tunable experimentally (for e.g., using Feshbach resonance in ultra-cold gases~\cite{chin2010feshbach,inouye1998observation, burchianti2018dual,ferioli2019collisions}). It is also potentially possible to design initial conditions and subsequently measure the time evolution~\cite{denschlag2000generating,burger2002generation,gajda1999optical}. Taking into consideration all the above, it is clear that MNLS is an ideal platform for our analysis. The MNLS under consideration is given by
\begin{equation}
 i \frac{\partial\psi_k}{\partial t}=-\frac{1}{2}\frac{\partial^2\psi_k}{\partial x^2}+\sum_{j=1}^{N} F_{kj}(|\psi_j|^2) \psi_k 
 \label{mnls}
\end{equation}
where $\psi_k (x,t)$ is a complex field (for the  $k$-th species) in one space and time dimension and the mass is set to $1$. Here, $N$ denotes the number of components or species (i.e., $k=1,2...N$). Therefore, Eq.~\ref{mnls} represents $N$ coupled complex partial differential equations (PDEs). This equation is a conservative system and for a suitable choice of Poisson bracket, it is in fact Hamiltonian. The elements of the interaction matrix $ F_{kj}(|\psi_j|^2)$ are functions of $ |\psi_j|^2$. The matrix structure we consider is given by $ F_{kj}(|\psi_j|^2)=\alpha_{kj}|\psi_j|^2  + \delta_{kj}G_j |\psi_j|^4$ where the coefficient $\alpha_{kj}$ defines both the self and cross couplings of cubic order and the coefficient $G_j$ specifies only self coupling of quintic type. This particular choice of  $F_{kj}$ is typical of several physical systems (for \textit{e.g.} cold atomic gases \cite{everitt2015observation,inouye1998observation,xi2016droplet,abdullaev2001stability,zakeri2018solitons,ramesh2010phase} and nonlinear optics \cite{pushkarov1996bright,skarka1997spatiotemporal,boudebs2003experimental}). In ultra-cold gases, the complex-valued function $\psi_k (x,t)$ represents the condensate wavefunction, the absolute-value squared of which is the density of atoms while its argument is the phase. On the other hand, in nonlinear optics, the complex field represents the pulse and the absolute-value squared is the intensity of light. In nonlinear optics, the role of time (in Eq.~\ref{mnls}) is played by an additional spatial $z$ direction \cite{agrawal2000nonlinear}.

Aside from certain exceptional choices for $F_{jk}$ \cite{manakov1974theory, forest2000nonfocusing,kevrekidis2015solitons}, Eq.~\ref{mnls} is in general \emph{not} integrable. Therefore for most choices of the couplings, one does not expect soliton solutions. One may still wonder if it is indeed possible to design special initial conditions $\psi_k (x,0)$ ($k=1,2...N$), which give rise to solutions of Eq.~\ref{mnls}, that are soliton-like in the sense that they bear as much resemblance as possible, to properties of true solitons of integrable models. In this paper, we ask whether there are localised solutions to Eq.~\ref{mnls} that propagate at almost constant speed? If so, do these localised solutions interact with one another with minimal scattering/radiation? As mentioned earlier, one suspects arbitrary initial conditions are subject to complex nonlinear dynamics, typically leading to breaking of profiles due to nonlinear and dispersive effects. Thus to answer such questions is not entirely trivial. In this work, we answer both questions in the affirmative and give a general recipe to find such initial conditions, for arbitrary coupling coefficients and background densities. In particular
\begin{itemize}


\item We obtain special soliton-like initial conditions for the original non-integrable MNLS (Eq.~\ref{mnls}) and show that the resulting time evolution bears a remarkable resemblance (Fig.~\ref{fig:intra_chiral} and Fig.~\ref{fig:inter_chiral}) to that of solitons of integrable models. 

\item We achieve the above by employing known multi-soliton ($M-$ soliton) solutions~\cite{hirota_2004} for the reduced chiral KdV equation (Eq.~\ref{derivedKdV}). These $M-$soliton solutions are used to construct special initial conditions for the original MNLS. As specific examples, we present analytical and numerical results for the two-soliton case ($M=2$) of the three component MNLS ($N=3$) but our method is valid for arbitrary $M$ and $N$. 

\item We demonstrate that even small perturbations of the optimal initial conditions invariably lead to radiation and subsequent failure of the robust evolution. This indicates that crude attempts to guess optimal initial conditions are doomed to fail (top panel of Fig.~\ref{fig:perturbed_linear}).

\item We manifestly establish that the initial condition so constructed exhibits dissimilar evolution under the original MNLS and the linearised version of MNLS. This establishes the fact that the dynamics is undoubtedly nonlinear (bottom panel of Fig.~\ref{fig:perturbed_linear}) and the robust evolution we find is far from trivial. 


\item We present a reduction to an effective uncoupled-system of KdV equations for the MNLS. Even though the KdV equations are uncoupled, the coefficients of each KdV encodes information of all the components in terms of base states (background fields) and coupling coefficients.

\item The MNLS-to-KdV reduction is based on a detailed spectral analysis of the linear version of MNLS and in the current work we address issues that were left open from an earlier work \cite{Swarup2020}. For example, we establish explicit expressions for eigenvectors which gives an analytical handle on deriving the coefficients of the effective chiral field theory. As a consequence we obtain a map from coupling coefficients and background densities of MNLS to parameters for the initial condition required to generate robustly propagating and stably interacting soliton-like solutions.

\item Although most of the results we present are for the case when the sound speeds of the underlying linear problem are distinct, we end the paper with a few statements regarding the case of repeated speeds. 


\end{itemize}

The paper is organised as follows. In Sec.~\ref{sec:SA} we recast MNLS Eq.~\ref{mnls} into suitable hydrodynamic variables (density and velocity fields). We then recap relevant results from an earlier work~\cite{Swarup2020} and present new results on the spectral analysis. In Sec.~\ref{sec:ECT} we  write down the effective chiral field-theory which is given by $2N$ (in general) \emph{decoupled} KdV equations ($N$ in each chiral sector) whose coefficients encode the information of the various components. $M-$ soliton solutions are discussed and the subsequent construction of optimal initial conditions is then explained. Sec.~\ref{sec:NS} is dedicated to numerical simulations. Comparisons between original (fully nonlinear) dynamics and the effective dynamics are made. We then make some important remarks regarding the interaction of these coherent solutions with one another. In Sec.~\ref{sec:CRSS} we briefly comment on the case of repeated sound speeds. More precisely, we give necessary and sufficient conditions to ensure distinct sound speeds thereby describing the parameter-space of validity for the results of Sec.~\ref{sec:ECT}.  When the sound speeds are not distinct, one remarkably obtains systems of coupled KdV equations the properties of which will be investigated in an upcoming work. We then summarise the results along with an outlook in Sec.~\ref{sec:C}. 

\section{Spectral Analysis}
\label{sec:SA}

We rewrite Eq.~\ref{mnls} in terms of density and velocity fields using a Madelung transformation~\cite{Madelung1929} $ \psi_k(x,t) = \sqrt{\rho_k(x,t)}e^{i \int_{0}^{x}v_k(x',t)dx'}$ to obtain 
%
       
\begin{eqnarray}
        \frac{\partial \rho_k}{\partial t}  =  - \frac{\partial}{\partial x} \big(\rho_k v_k\big),\quad 
        \frac{\partial v_k}{\partial t} = -\frac{\partial}{\partial x}\bigg[\sum_{j=1}^{N}\alpha_{kj}\rho_j + \frac{v_k^2}{2} + G_{k}\rho_k^2- \frac{1}{2}\frac{\partial_x^2\sqrt{\rho_k}}{\sqrt{\rho_k}}\bigg]. \nonumber \\ \label{mcqv}
\end{eqnarray}       
       
We then linearise the above problem about a base state ($\rho_{0k},0 $) with non-zero background densities and zero background velocity. We define the set of perturbed fields ($ \delta\rho_{k}, \delta v_{k} $) such that $\rho_k=\rho_{0k}+\epsilon^2\delta\rho_k(\epsilon x, \epsilon t)$ and $v_k=\epsilon^2 \delta v_k(\epsilon x,\epsilon t)$ where $\epsilon$ is a small parameter. We substitute the form of the perturbation into Eq.~\ref{mcqv} and drop terms of $\mathcal{O}(\epsilon^2)$ and higher to obtain the linear equation

\begin{equation}
		\partial_t
		\begin{pmatrix}
		\delta\rho
		\\
		\delta v
		\end{pmatrix} =-\partial_x 
		\mathcal A
		\begin{pmatrix}
		\delta\rho\\
		\delta v
		\end{pmatrix} \text{ where }  \mathcal A = \left( \begin{array}{cc} \mathbf{0}_{N\times N} & \rho \\ \tilde{\alpha} & \mathbf{0}_{N\times N} \end{array} \right).
\label{eq:matA}
\end{equation}
	 Note $\rho$  is an $N\times N$ diagonal matrix with strictly positive elements $\rho_{0k}$ (the background densities) and $\delta \rho\,,\delta v$ are the $N\times 1$ column vectors for the perturbations in density and velocity. 
	The matrix $\tilde{\alpha}$ is given by~\cite{Swarup2020}
\begin{align}
\label{eq:tildealpha}
 \tilde{\alpha}_{ij}= \left\{\begin{array}{ll} \tilde{g}_{i}, & i=j, \\ h, & i\neq j. \end{array} \right. \text{ where } \tilde{g}_{i}=g_i +2G_i \rho_{0i}
\end{align}
with $g_i$ being the diagonal elements of the $\alpha$ matrix. All elements of the  $\tilde{\alpha}$ matrix (Eq.~\ref{eq:tildealpha}) are strictly positive. The spectral analysis of the $2N \times 2N$ matrix $\mathcal A$ (Eq.~\ref{eq:matA}) can be accomplished by a spectral analysis of the $N\times N$ matrix $\tilde{\alpha}\rho$ \cite{Swarup2020}. Indeed, under the assumption that $\tilde{\alpha}$ is symmetric positive definite (for $\tilde{\alpha}$ as defined above this is true when $h<\min{\tilde{g}_i}$~\cite{Swarup2020}), all eigenvalues of $\tilde{\alpha}\rho$ are real and positive which we denote by $\lambda^2$. Additionally, the matrix $\tilde{\alpha}\rho$ is diagonalisable. Equivalently the algebraic multiplicity (degeneracy of eigenvalue) and the geometric multiplicity (dimension of the span of all eigenvectors associated with the degenerate eigenvalue) of every eigenvalue of $\tilde{\alpha} \rho$ are equal. And lastly, to each eigenvector of $\tilde{\alpha}\rho$ there corresponds an independent eigenvector of $\mathcal A$. Hence $\mathcal A$ too is diagonalisable and explicit expressions for its eigenvectors can be stated in terms of eigenvectors of $\tilde{\alpha} \rho$. We will discuss the case when an eigenvalue of $\tilde{\alpha}\rho$ has multiplicity $m=1$, \textit{i.e.} the case of a simple eigenvalue. The case of repeated eigenvalues (which corresponds to the case of repeated or non-unique sound speeds) and necessary and sufficient conditions for that to happen is discussed later (see Sec.~\ref{sec:CRSS}). 

Here we provide an explicit form for the eigenvector of $\tilde{\alpha}\rho$ (and hence for $\mathcal A$) when $\lambda^2$ is a simple eigenvalue ($m=1$). We assume we know the value of this eigenvalue and then construct the associated eigenvector. If $\lambda^2$ is a simple eigenvalue of $\rho\tilde{\alpha}$ with eigenvector $q$ then $\rho\tilde{\alpha} q = \lambda^2 q \Rightarrow (\rho\tilde{\alpha} - \lambda^2)q = 0$. Thus to determine the eigenvector of $\rho\tilde{\alpha}$ it suffices to determine the null-vector of $\rho\tilde{\alpha}-\lambda^2$. 

It so happens that one can readily write down eigenvectors for all simple eigenvalues if one can write down the determinant of the matrix $\rho\tilde{\alpha}-\lambda^2$. To explain this point we take an important detour. Suppose $B$ is a $3\times 3$ matrix of the particular form
\begin{equation}
B = \begin{pmatrix} a_1 & a_2 & a_3 \\ b_1 & b_2 & b_3 \\ c_1 & c_2 & c_3 \end{pmatrix} = \begin{pmatrix} a^T \\ b^T \\ c^T \end{pmatrix}
\end{equation}
with zero determinant and hence there exists a non-trivial null-vector. Suppose zero is a simple eigenvalue of $B$ and hence the null-space is spanned by a single vector (up to scaling). Using the standard cross product in $3-$dimensions, the vector $b\times c$ is orthogonal to vectors $b$ and $c$. On the other hand, the determinant of the singular matrix $B$ is given by $\mbox{det}B = a_1(b_2c_3-b_3c_2) - a_2(b_1c_3-c_1b_3) + a_3(b_1c_2-c_1b_2) = a\cdot (b\times c) = 0$,
and hence $b\times c$ is orthogonal to $a,b$ and $c$. Since $b\times c$ is orthogonal to every row of $B$, we have determined a null-vector for the matrix $B$. This implies we have determined the eigenvector for the null-space of $B$. If $B=A-\lambda$ where $\lambda$ is a simple eigenvalue of $A$, then we have determined the eigenvector of $A$ associated with eigenvalue $\lambda$. Recall $a\cdot (b\times c) = c\cdot (a\times b) = b\cdot(c\times a)$. As a consequence, $b\times c$, $a\times b$ and $c\times a$ are all equally valid expressions for the null-vector of $B$. However as zero is a simple eigenvalue of $B$, the null-space is spanned by a single vector and the seemingly different expressions are in fact linearly dependent: they are equal up to scaling. 

The idea explained above easily generalises into higher dimensions. Indeed the determinant of the same $3\times 3$ matrix $B$ may also be written as $a_ib_jc_k \varepsilon_{ijk}$ where $\varepsilon_{ijk}$ is the fully anti-symmetric alternating tensor which equals $+1$ when $i,j,k$ is an even permutation of $1,2,3$; equals $-1$ when $i,j,k$ is an odd permutation of $1,2,3$ and zero otherwise (Einstein summation is assumed). Thus an expression for the null-vector is $b_jc_k\varepsilon_{ijk}$.  This particular representation extends to larger matrices in a straightforward manner. Suppose $B$ were an $N\times N$ matrix with $b^{(i)}$, $i=1,2,\ldots,N$, the rows of $B$. Then 
$\mathrm{det } B = b^{(1)}_{i_1}\cdots b^{(N)}_{i_N}\varepsilon_{i_1\cdots i_N}$
where $\varepsilon_{i_1\cdots i_N}$ is the fully anti-symmetric tensor in $N$ symbols. If $\mbox{det }B=0$ then we readily have an expression for the null-vector which is given by $b^{(2)}_{i_2}\cdots b^{(N)}_{i_N}\varepsilon_{i_1\cdots i_N}$. The main takeaway from the preceeding discussion is that the algebraic expression for the determinant of a matrix, when suitably interpreted, contains the elements of the null-vector (equivalently, the eigenvector). One only has to `recognise' the determinant as a dot-product between any row of the matrix with some other vector. The other vector in question, is the eigenvector. Note this statement does not assume any special structure for the matrix.

We now return to the problem at hand.
Note that $\rho\tilde{\alpha} - \lambda^2 = h\rho X_N \Rightarrow \mbox{det }(\rho \tilde{\alpha}-\lambda^2) = (\mbox{det}X_N) \prod_{i=1}^N(\rho_{0i} h)$ 
where the elements of the $N\times N$ matrix $X_N$ are given by 
\begin{equation}
\label{eq:XNmat}
(X_N)_{ij} = \left\{ \begin{array}{ll} \gamma_i,& i=j,\\ 1, & i\neq j, \end{array}\right.\quad \gamma_i = \frac{\rho_{0i} \tilde{g}_i-\lambda^2}{\rho_{0i} h}.
\end{equation}
Recall, in our previous work~\cite{Swarup2020}, we showed that $\mbox{det }X_N$ satisfied a recursion relation. Let us define the notation $[\gamma_1 \cdots \gamma_N] \equiv \mbox{det }X_N$. Then the recursion relation is
\begin{equation}
[\gamma_1 \cdots \gamma_N] = \gamma_N[\gamma_1\cdots \gamma_{N-1}] - \sum_{k=1}^{N-1}[\gamma_1\cdots \gamma_{k-1}\: 1 \: \gamma_{k+1}\cdots\gamma_{N-1}].
\end{equation}
Here $[\gamma_1\cdots \gamma_{N-1}]$ is the determinant of the $(N-1)\times(N-1)$ matrix $X_{N-1}$ with $\gamma_i,\: i=1,\ldots,N-1$ along the diagonal. Likewise the term in the summation is the determinant of $X_{N-1}$ but with $1$ for $k$-th element along the diagonal. This implies 
\begin{align} 
\mbox{det}(\rho\tilde{\alpha}-\lambda^2) =  \tilde\gamma_N[\tilde\gamma_1\cdots \tilde\gamma_{N-1}]_\rho - \rho_{0N} h \sum_{k=1}^{N-1}[\tilde\gamma_1\cdots \tilde\gamma_{k-1} \: \rho_{0k} h\: \tilde\gamma_{k+1}\cdots\tilde\gamma_{N-1}]_\rho 
\label{eq:sub_rho1}
\end{align}
where
\begin{equation}
[\tilde\gamma_1\cdots \tilde\gamma_{N-1}]_\rho = \mbox{det}\tilde X_{N-1},\quad (\tilde X_{N-1})_{ij} = \left\{ \begin{array}{ll} \tilde\gamma_i,& i=j,\\ h \rho_{0i}, & i\neq j, \end{array}\right.\text{ and }\quad \tilde\gamma_i = \rho_{0i} \tilde{g}_i-\lambda^2.
\label{eq:sub_rho2}
\end{equation}
The subscript $\rho$ in Eq.~\ref{eq:sub_rho1} and  Eq.~\ref{eq:sub_rho2} indicates that the concerned matrices contain $h \rho_{0i}$ as off-diagonal elements. The elements within the square brackets denote the diagonal elements of $\tilde X_{N-1}$. It is easy to note that the $N-$dimensional vector $\big(\rho_{0N} h,  \rho_{0N} h \cdots \rho_{0N} h, \quad (\rho_{0N}g_N-\lambda^2)\: \big)^T$ is the last row of $(\rho \tilde{\alpha}-\lambda^2)$. Moreover we see that Eq.~\ref{eq:sub_rho1} expresses the determinant of $\tilde X_N$ as a dot-product of this vector (a row of $\tilde X_N$) with some other vector. We write out this vector as 
\begin{equation}
p \equiv 
\begin{pmatrix} 
-[h\rho_{01}  \: \tilde\gamma_2 \cdots \tilde \gamma_{N-1} ]_\rho\\
-[\tilde \gamma_1 \: h\rho_{02}  \: \tilde\gamma_3 \cdots \tilde \gamma_{N-1} ]_\rho\\
\vdots \\
-[\tilde\gamma_1 \cdots \tilde\gamma_{N-2} \: h\rho_{0N-1} ]_\rho \\ 
[\tilde\gamma_1\cdots \tilde\gamma_{N-1}]_\rho
\end{pmatrix}.
\label{eq:eig}
\end{equation}
Hence Eq.~\ref{eq:eig} is an explicit representation for the simple eigenvector associated with eigenvalue $\lambda^2$ for the matrix $\rho \tilde{\alpha}$. From our previous results \cite{Swarup2020}, we have explicit formulae for elements of the eigenvector. Indeed
\begin{align}
[\tilde\gamma_1\cdots \tilde\gamma_{N-1}]_\rho &= [\gamma_1\cdots \gamma_{N-1}] \prod_{i=1}^{N-1}(h\rho_{0i})\\
[\tilde\gamma_1\cdots \tilde\gamma_{k-1} \: \rho_{0k} h\: \tilde\gamma_{k+1}\cdots\tilde\gamma_{N-1}]_\rho &= [\gamma_1\cdots  \gamma_{k-1}\: 1 \: \gamma_{k+1}\cdots \gamma_{N-1}]\: \prod_{i=1}^{N-1}(h\rho_{0i})
\end{align}
where 
\begin{align}
[\gamma_1\cdots \gamma_{N}]= \mbox{Sym}(\gamma_i,N) + \sum_{k=2}^N (-1)^{k-1}(k-1)\mbox{Sym}(\gamma_i,N-k),\quad \gamma_i = \frac{\rho_{0i} \tilde{g}_i-\lambda^2}{h\rho_{0i}}.
\end{align}
At this point we present some illustrative examples.  Let $\lambda^2$ be a known simple eigenvalue of the three component ($N=3$) case. Then the matrix
\begin{equation}
\rho \tilde{\alpha} - \lambda^2 = \begin{pmatrix} \tilde\gamma_1 & h\rho_{01} & h\rho_{01} \\ h\rho_{02} & \tilde \gamma_2 & h\rho_{02} \\ h\rho_{03} & h\rho_{03} & \tilde\gamma_3\end{pmatrix}, \quad \tilde\gamma_i = \rho_
{0i}\tilde{g}_i-\lambda^2
\end{equation}
has vanishing determinant and 
\begin{align}
\begin{pmatrix}
     -[h\rho_{01}, \tilde\gamma_2]_{\rho}\\
     -[\tilde\gamma_1, h\rho_{02}]_\rho\\
     [\tilde\gamma_1, \tilde\gamma_2]_\rho
\end{pmatrix} = 
\begin{pmatrix}
     -\tilde\gamma_2h\rho_{01} + \rho_{01}\rho_{02}h^2\\
     -\tilde\gamma_1h\rho_{02} + \rho_{01}\rho_{02}h^2\\
     \tilde \gamma_1 \tilde \gamma_2 - \rho_{01}\rho_{02}h^2
\end{pmatrix}
\end{align}
is the associated eigenvector of $\rho\tilde{\alpha}$. Indeed one can compute the product of the matrix  $(\rho \tilde{\alpha} - \lambda^2)$ with the vector given above to obtain the trivial vector if and only if $\mathrm{det}(\rho \tilde{\alpha} - \lambda^2)=0$. 

Likewise for $N=4$ we explicitly have
\begin{align}
\begin{pmatrix}
    -h\rho_{01}\tilde\gamma_2\tilde\gamma_3 + (\rho_{03}\tilde\gamma_2+\rho_{02}\tilde\gamma_3)\rho_{01}h^2 - \rho_{01}\rho_{02}\rho_{03}h^3\\
    -\tilde\gamma_1 h\rho_{02}\tilde\gamma_3 + (\rho_{03}\tilde\gamma_1+\rho_{01}\tilde\gamma_3)\rho_{02}h^2 - \rho_{01}\rho_{02}\rho_{03}h^3\\
    -\tilde\gamma_1\tilde\gamma_2 h\rho_{03} + (\rho_{02}\tilde\gamma_1+\rho_{01}\tilde\gamma_2)\rho_{03} h^2 - \rho_{01}\rho_{02}\rho_{03}h^3\\
    \tilde\gamma_1\tilde\gamma_2\tilde\gamma_3 - (\rho_{02}\rho_{03}\tilde\gamma_1+\rho_{01}\rho_{03}\tilde\gamma_2+\rho_{01}\rho_{02}\tilde\gamma_3)h^2 + 2\rho_{01}\rho_{02}\rho_{03}h^3
\end{pmatrix}
\end{align}
and the general case (for any $N$) is given by Eq.~\ref{eq:eig} which  allows us to explicitly write an eigenvector $p$ of $\rho\tilde{\alpha}$ with simple eigenvalue $\lambda^2$. We conclude by noting that $[p \quad \pm \lambda \rho^{-1} p]^T$ is an eigenvector of $\mathcal A$ (Eq.~\ref{eq:matA}) with eigenvalue $\pm \lambda$ and $q\equiv \rho^{-1}p$ is then the eigenvector of $\tilde{\alpha}\rho$.

Before we move to the next section which deals with chiral theory, we introduce some matrices which are of use there. Let $Q$ be the $N \times N$ matrix of eigenvectors of  $\tilde{\alpha} \rho$ (i.e., $Q$ represents the matrix with columns $q_{i}$ where $q_{i}$ is the $i-$th eigenvector of  $\tilde{\alpha}\rho$). Let $\Lambda$ be the $N \times N$ diagonal matrix containing positive eigenvalues ($\lambda$) and let $L = Q^{T} \rho Q$. We then have the following $2N \times 2N$ matrix $V$ whose columns are eigenvectors of $\mathcal A$ (Eq.~\ref{eq:matA})
 
 \begin{equation}
 \label{eq:vmat}
 V = \left(\begin{array}{cc} \rho Q \Lambda^{-1} & - \rho Q \Lambda^{-1} \\ Q & Q \end{array}\right), \quad\quad (V^{-1})^T = \frac 1 2 \left(\begin{array}{cc} Q L^{-1} \Lambda & - QL^{-1}\Lambda \\ \rho QL^{-1} & \rho QL^{-1} \end{array}\right).
 \end{equation} 
A major step forward in comparison to the earlier work \cite{Swarup2020} is that we now can analytically compute $Q$ and therefore $V$ (Eq.~\ref{eq:vmat}). The relevant input is the coupling coefficients and background densities, once the eigenvalue (sound speed) has been determined.



\section{Effective chiral theory and optimal initial conditions}
\label{sec:ECT}

This section is dedicated to writing down the effective chiral theory (Sec.~\ref{sec:ECT1}) and then discussing our prescription to generate optimal initial conditions (Sec.~\ref{sec:ECT2}). 

\subsection{Effective chiral theory}
\label{sec:ECT1}

Recall the original MNLS Eq.~\ref{mnls} is a system of nonlinear coupled PDEs of $N$ complex functions which can be equivalently written as a system of $2N$ PDEs for real fields (Eq.~\ref{mcqv}). Here the central question is whether we can make a chiral reduction \cite{Swarup2020} to $N$ chiral equations ($N$ in each chiral sector). It turns out that we can indeed reduce the system into a total of $2N$ uncoupled KdV equations (when all eigenvalues of $\mathcal{A}$ are simple, \textit{i.e.} distinct sound speeds)
\begin{align} \label{derivedKdV}
 \partial_{\tau}f_j+K_{NL}f_j\partial_{\xi} f_j+K_{DS}\partial_{\xi}^{3}f_j=0,\quad j = 1,2...N\,\, (\text{for each chiral sector}).
\end{align}
We now describe the ingredients of Eq.~\ref{derivedKdV}.  Every KdV equation in Eq.~\ref{derivedKdV} is for a particular eigenvalue $\lambda_j$ (speed of sound; for the subsequent discussion we fix a specific $j$ value). Using the small parameter $\epsilon$, the new (stretched) space-time variables ($\xi$ and $\tau$) used in Eq.~\ref{derivedKdV} are given in terms of the old space-time variables ($x$ and $t$) by the expressions $\xi = \epsilon x - \epsilon \lambda_j t$ and $\tau = \epsilon^3 t$. 	
In Eq.~\ref{derivedKdV}, $f_j(\xi, \tau)$ are chiral fields in stretched variables and are related to the original fields by  
\begin{align}
		\begin{pmatrix}
		\vec{\rho}\\
		\vec{v}
		\end{pmatrix} 
		= 
		\begin{pmatrix}
		\vec{\rho}_0 \\
		0
		\end{pmatrix}
		+ \epsilon^2  f_j(\xi, \tau)V_j+ \mathcal{O}(\epsilon^4),\qquad  V_j = 
		\begin{pmatrix}
		\rho q_{j}/\lambda_j\\
		q_{j}
		\end{pmatrix}. 
		\label{eq:hydro_f}
	\end{align}
%
Here $\vec{\rho}$ and $\vec{v}$ denote the set of density fields $\{\rho_k(x,t)\}$ and velocity fields $\{v_k(x,t)\}$. 
Furthermore $\vec{\rho}_0$ is the set of background densities $\{\rho_{0k}\}$  and as before $\rho$ denotes the $N\times N$ diagonal matrix with $\rho_{0k}$ along the diagonal.
$V_j$ denotes the $j-$th column of the $2N \times 2N$ matrix $V$ (Eq.~\ref{eq:vmat}). 
Looking at Eq.~\ref{eq:vmat}, it is easy to note that the  $j-$th column of $V$ is given by the elements $\rho_{0i} q_{j}^{(i)}/\lambda_j$ for $i=1,2,..N$ and $q_{j}^{(i)}$ for $i=N+1,N+2,..2N$ where $q_{j}^{(i)}$ is the $i-$th element of the $N\times 1$ column vector $q_j$. 
Note $q_{j}$ is the eigenvector of $\tilde\alpha\rho$ corresponding to eigenvalue $\lambda_j^2$ and can be readily extracted from Eq.~\ref{eq:eig} (along with the fact that $q_j\equiv \rho^{-1}p_j$). It remains to discuss the nonlinear and dispersive coefficients, $K_{NL}$ and $K_{DS}$ respectively. For this purpose, we introduce a $2N \times 1$ vector $\mathcal {N}$ given by~\cite{Swarup2020}		
\begin{align}
\mathcal{N} &= f_j \partial_\xi {f_j}\: \mathcal{N}_A + \partial_\xi^3 f_j\: \mathcal{N}_B  \nonumber \\
&=  
f_j \partial_\xi {f_j}  
\begin{pmatrix}
\dfrac{2 \rho \, (q_j \circ q_j)}{\lambda_j }\\
		(q_j \circ q_j ) + \dfrac{2}{\lambda_j^2}\: G\circ (\rho q_j) \circ (\rho q_j)
\end{pmatrix} +
\partial_\xi^3 f_j 
\begin{pmatrix}
0\\
		-\dfrac{q_j}{4\lambda_j}  
\end{pmatrix}
\label{eq:Nmat}
\end{align}


\noindent where $``\circ"$ in Eq.~\ref{eq:Nmat} denotes the element-wise multiplication of $N \times 1$ vectors. Eq.~\ref{eq:Nmat} says that $\mathcal{N}$ is a $2N \times 1$ column vector with elements given by $ f_j \partial_\xi {f_j}  \dfrac{2\rho_{0i}\big(q_j^{(i)}\big)^2}{\lambda_j} $ for  $i=1,2,..N$ and $f_j \partial_\xi f_{j} \left(  \big(q_j^{(i)}\big)^2  + \dfrac{2}{\lambda_j^2} G_i  \big(\rho_{0i}q_j^{(i)}\big)^2   \right)  -\partial_\xi^3 f_j \dfrac{q_j^{(i)}}{4\lambda_j} $ for $i=N+1,N+2,..2N$. The KdV equation (Eq.~\ref{derivedKdV}) can be equivalently written as $\partial_\tau f_j +  \left\langle (V^{-1})^T_j  \mid \mathcal{N}\right\rangle=0$ where $(V^{-1})^T_j $ is the $j-$th column of the $2N \times 2N$ matrix $(V^{-1})^T$ given in Eq.~\ref{eq:vmat}. More precisely, 
\begin{equation}
(V^{-1})^T_j = \frac{1}{2}\begin{pmatrix}  q_j\lambda_j/l_j  \\ \rho q_j/l_j \end{pmatrix}
\end{equation}
where $l_j$ are the diagonal elements of the matrix $L$ in Eq.~\ref{eq:vmat} (in fact $L$ is a diagonal matrix~\cite{Swarup2020}). It is easy to see that $l_j = \langle q_j \mid \rho q_j \rangle$.
We therefore finally get the coefficients $K_{NL}$ and $K_{DS}$ 

\begin{align}
\label{eq:KNL}
K_{NL} &=  \left\langle (V^{-1})^T_j  \mid \mathcal{N}_A\right\rangle  = \frac{3}{2l_j} \sum_{i=1}^{N}  \rho_{0i}\, \big(q_j^{(i)}\big)^3 +
\frac{1}{\lambda_j^2\: l_j } \sum_{i=1}^{N} G_i\, \big(\rho_{0i}\: q_j^{(i)}\big)^3
  \\
K_{DS} &=  \left\langle (V^{-1})^T_j  \mid \mathcal{N}_B\right\rangle  = - \frac{1}{8\lambda_j l_j } \sum_{i=1}^{N}  \rho_{0i}\, \big(q_j^{(i)}\big)^2  .
\label{eq:KDS}
\end{align}
And finally we emphasize even though the KdV equations (Eq.~\ref{derivedKdV}) are uncoupled, the coefficients $K_{NL}$ and $K_{DS}$ for each KdV equation depends on the background fields and coupling coefficients of all the components.

\subsection{Multi-soliton solutions of the Chiral PDE and Optimal Initial condition for $\psi_k$}
\label{sec:ECT2}

The $M-$soliton solutions of the integrable Eq.~\ref{derivedKdV} are given by~\cite{hirota_2004}
\begin{eqnarray}
\label{nsolitonkdvoriginal}
f_j(\xi,\tau) & = \frac{12 K_{DS}}{K_{NL}} \partial_{\xi}^2\log \bigg( \sum_{\{\sigma_a, \sigma_b\ \}\in \{0,1 \}^2} e^{W[\sigma_a,\sigma_b ]} \bigg)
\label{eq:msol}
\end{eqnarray}

with
\begin{equation}
W[\sigma_a,\sigma_b ]= {\sum_{a=1}^{M}\sigma_{a}\eta_{a}(\xi,\tau)+\sum_{\substack{a,b =1 \\ a< b}}^{M}\log\left(\frac{k_a -k_b}{k_a + k_b}\right)^2\sigma_{a}\sigma_{b}}
\end{equation}
and
\begin{equation}
\eta_a(\xi,\tau) =  \sqrt{\frac{K_{NL}}{6 K_{DS}}}\left( k_a \xi -  \frac{K_{NL}}{6}k_a^3\tau\right) - \eta_a^{(0)} . 
\label{eq:msol_ex}
\end{equation}

\noindent Here,  $ \sigma_{a}=0,1 $ (where the subscript `$a$' runs from $1$ to $M$) and $\sum $ in Eq.~\ref{eq:msol} includes a sum over all possible pairs $\{\sigma_a, \sigma_b \}\in \big\{ \{0,0\},  \{1,0\},  \{0,1\},  \{1,1\} \big\}$.  The soliton parameters are given by $k_a$ and $\eta_{a}^{(0)}$ and these parameters are chosen such that $\eta_a(\xi,\tau)$ is real. Eq.~\ref{eq:msol} is used to construct the optimal initial conditions for the original MNLS Eq.~\ref{mnls}. This is done by (i) moving back from the stretched variables ($\xi,\tau$) to original variables ($x,t$) by fixing a particular small $\epsilon$, (ii) using Eq.~\ref{eq:hydro_f} to construct the initial density and velocity fields, and (iii) performing an inverse Madelung transformation to get the optimal $\psi_k (x,0)$ which can then be employed as initial conditions for Eq.~\ref{mnls}. We emphasize again, aside from the soliton parameters which are free variables, the physically relevant information needed to construct these initial conditions is just the set of background densities and coupling coefficients. In the next section, we will discuss numerical experiments that were performed following the above procedure and the remarkable consequence of soliton-like interaction.

\section{Numerical Simulations}
\label{sec:NS}

\begin{figure}
    \centering
    \includegraphics[width=\textwidth]{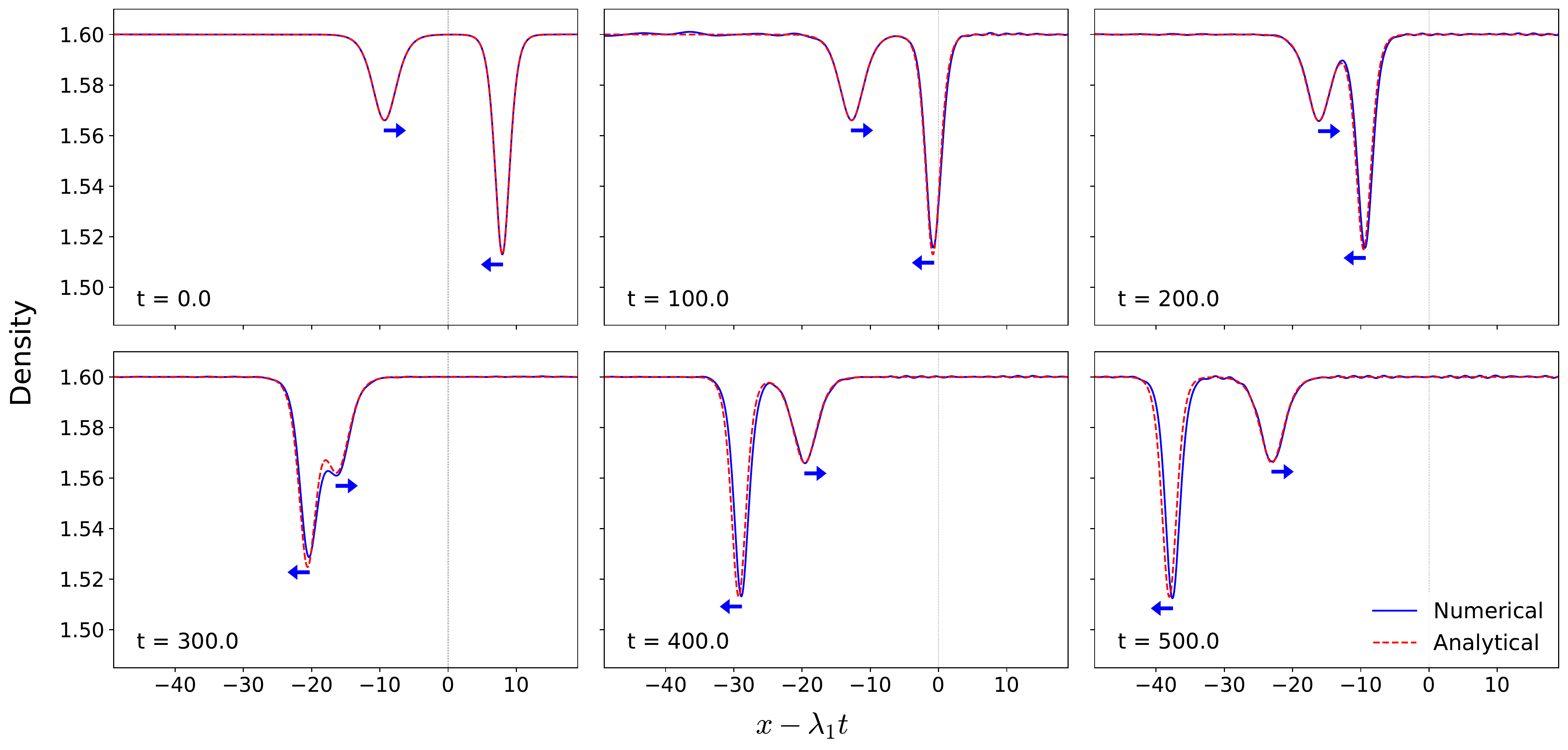}
    \vspace{4mm}
    \includegraphics[width=0.54\textwidth]{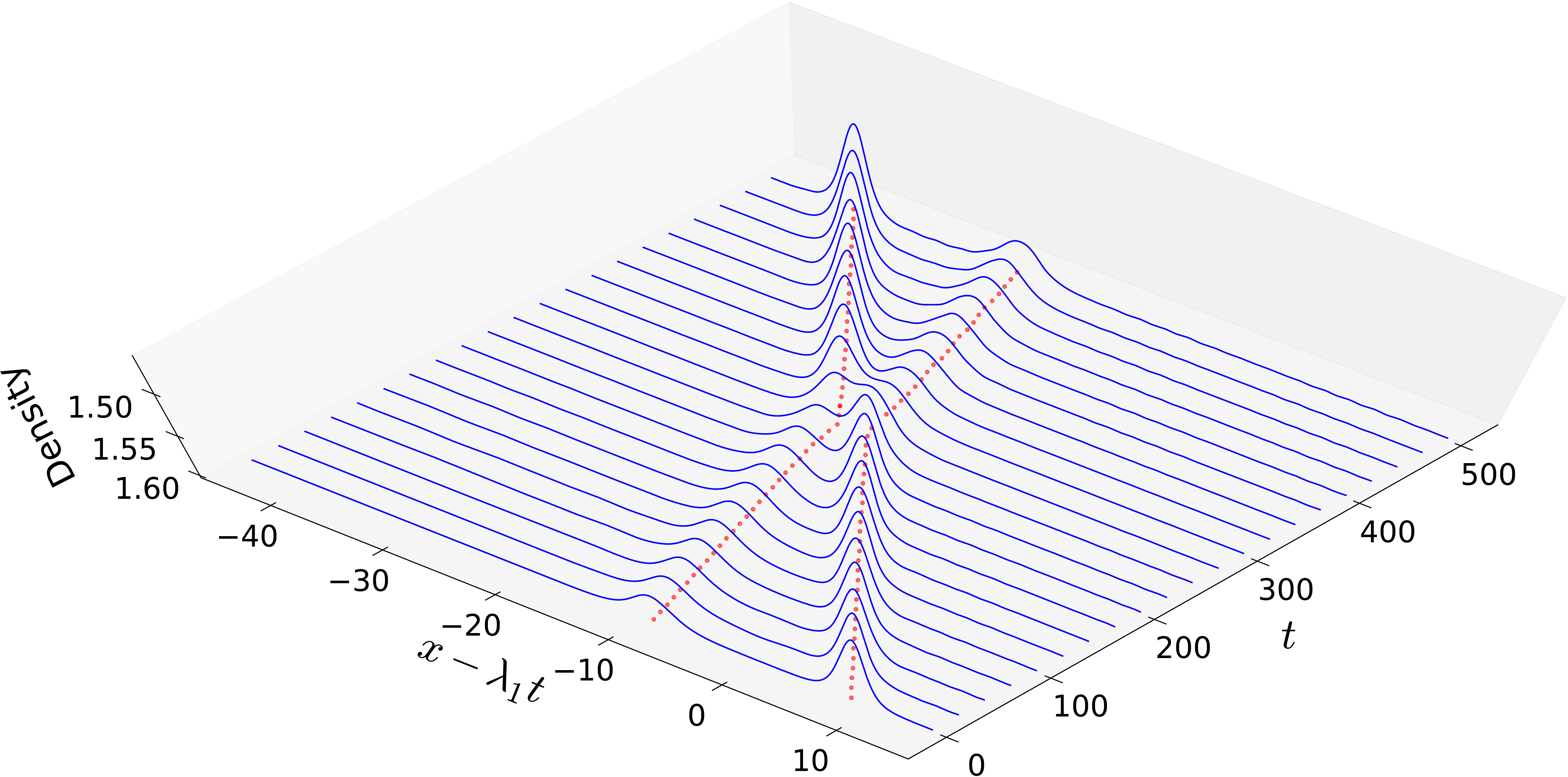}
    \includegraphics[width=0.44\textwidth]{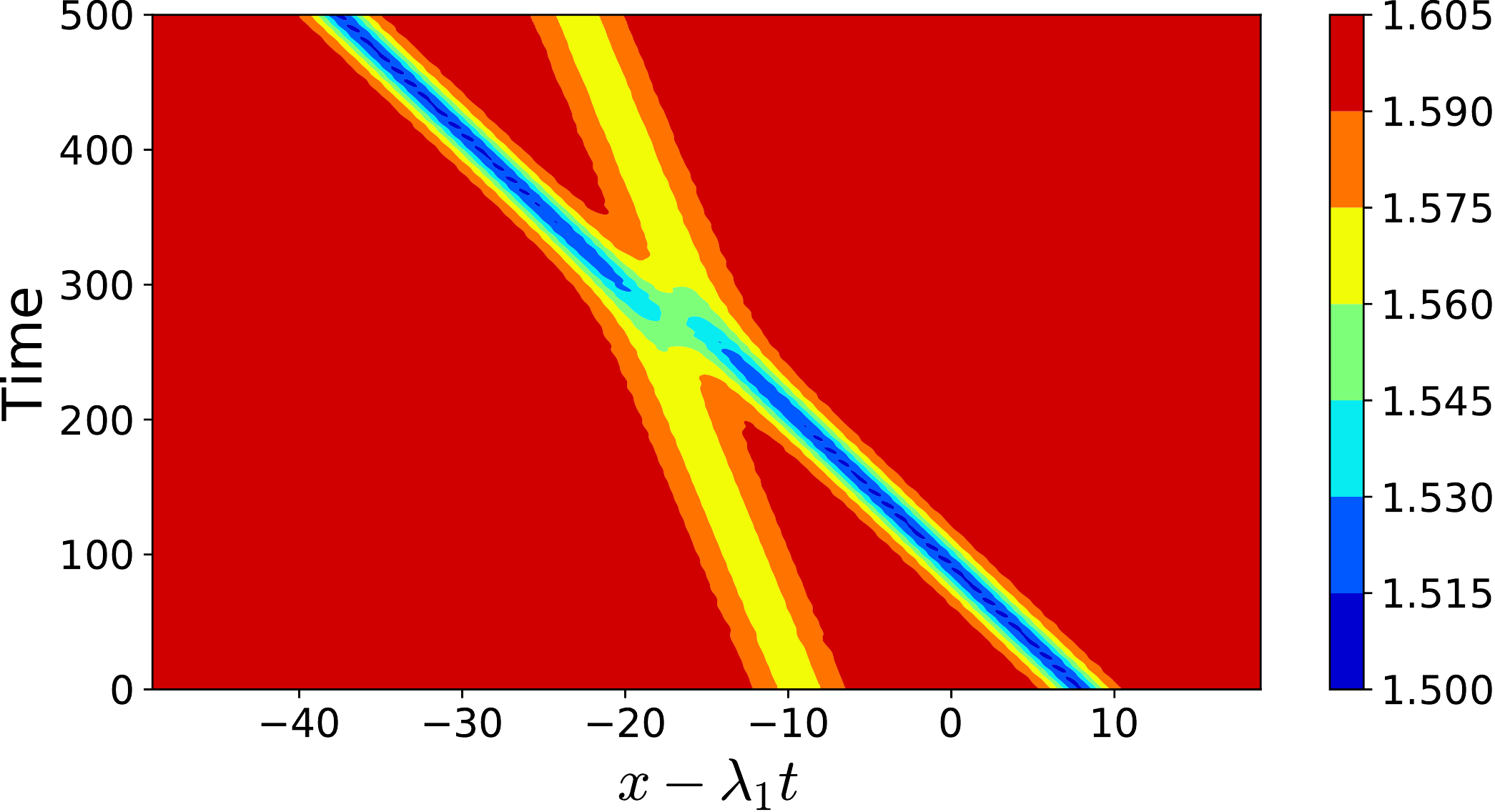}
    \caption{\footnotesize (colour on-line) (Top) Comparison of the numerical evolution of density of the original dynamics (Eq.~\ref{mnls}) and the chiral dynamics (Eq.~\ref{derivedKdV}). 
    The arrows indicates the direction of the two soliton peaks. 
    (Bottom) Time evolution of the two soliton-like peaks with the original dynamics  (Eq.~\ref{mnls}). Bottom left shows a three dimensional plot (along with the peak positions marked in red) and the bottom right shows the contour plot of the same figure. The contour plot shows how the soliton-like profiles essentially pass through each other. 
    For all the plots, we choose the following values for the parameters: $g_1=1, \; g_2=1.2,\; g_3=1.3,\;h=0.5,\; \epsilon=0.3,\; \rho_{01}=1.6,\; \rho_{02}=1.8,\;\rho_{03}=1.4,\; G_1=0.8,\;G_2=0.6,\;G_3=0.8$, $dx= 0.046875$ and $dt= 5.0\times10^{-4}$. The soliton parameters are chosen as $k_1=1.0,\;k_2=1.6,\;\eta_{1}^{0}=-8.0,\;\eta_{2}^{0}=8.0$. 
    The plots show time evolution of the first component of the three component system ($N=3$) evolved using the finite difference scheme given by Eq.~\ref{discretizedmnls}.} 
    \label{fig:intra_chiral}
\end{figure}


This section is devoted to numerical simulations and has three subsections. In Sec.~\ref{sec:intra} we compare the time evolution between original 
dynamics (Eq.~\ref{mnls}) and the effective chiral dynamics (Eq.~\ref{derivedKdV}) for the optimally chosen initial condition. In particular, without loss of generality, we will discuss the two-soliton case ($M=2$) for the three component MNLS ($N=3$). Remarkable soliton-like properties are exhibited. In Sec.~\ref{sec:adhoc_linear} we demonstrate that (i) ad hoc guesses for optimal initial condition are doomed to fail and (ii) the dynamics of our optimal initial condition (obtained by the protocol discussed in previous section) are definitely nonlinear. In Sec.~\ref{sec:inter} we present an interesting case of bidirectional motion, which strictly speaking goes beyond the formalism discussed in this paper, but nevertheless displays unexpected soliton-like behaviour. 

Before we discuss our numerical results, we briefly explain the method involved \cite{TAHA1984203}. Eq. \ref{mnls} is discretised using a central difference scheme in space and time, and its finite difference representation is given by

\begin{align}
    i\: \frac{\psi_k^{m,t+\Delta t} - \psi_k^{m,t-\Delta t}}{2 \Delta t} = \frac{1}{2}\frac{\psi_k^{m+1,t}-2\psi_k^{m,t}+\psi_k^{m-1,t}}{\Delta x^2} + \sum_{j=1}^{N} F_{kj}(|\psi_j^{m,t}|^2) \psi_k^{m,t} 
    \label{discretizedmnls}
\end{align}
where $\psi_k^{m,t}$ is the value of $\psi$ for $k^{th}$ component at grid-point $m$ and at time-step $t$. The method is linearly (dropping the nonlinear terms) stable for $\Delta t / (\Delta x)^2 \leq 1/4$ with truncation error in the scheme of the order $\big(\mathcal{O}(\Delta t)^2 + \mathcal{O}(\Delta x)^2\big)$. No-flux boundary condition $\partial{\psi_k}/\partial x = 0$ is used at the boundary points and the domain is large enough to ensure that there are no boundary effects. The fourth-order Runge-Kutta scheme was used for the first time step and Eq.~\ref{discretizedmnls} was used for the subsequent time steps. Similarly, the hydrodynamic version (Eq.~\ref{mcqv}) was also numerically solved by finite difference.

\subsection{Comparison between Original dynamics and the effective chiral dynamics: Soliton-like behaviour}
\label{sec:intra}
Here we compare the time evolution of the original dynamics and the reduced chiral dynamics. 
As an example  we explicitly discuss the three component ($N=3$) MNLS case admitting a two-soliton ($M=2$) solution for the chiral equation. Using the effective chiral theory, MNLS reduces to six distinct KdV equations, each corresponding to a distinct eigenvalue $\lambda_j$ ($j=1, 2, 3$ for one chiral sector and $j=4, 5, 6$ for the other chiral sector). We pick one such KdV (Eq.~\ref{derivedKdV}) corresponding to a distinct eigenvalue, say  $\lambda_1$ (i.e., $j=1$). A two-soliton initial condition solution $f_1(\xi,\tau=0)$ is obtained using Eq.~\ref{nsolitonkdvoriginal} (by substituting $M=2$). 
We then get the initial condition for density  $\{\rho_1(x,0),\rho_2(x,0),\rho_3(x,0)\}$ and velocity, $\{v_1(x,0),v_2(x,0),v_3(x,0)\}$ using Eq.~\ref{eq:hydro_f}. Once we have this,  we employ the Madelung transformation to get $\{ \psi_1 (x,0), \psi_2 (x,0), \psi_3 (x,0)\}$ given as
\begin{equation}
\label{eq:psi_init}
    \psi_k(x,0) = \sqrt{\rho_{0k} + \epsilon^2 \frac{\rho_{0k} q_1^{(k)}}{\lambda_1} f_1(\epsilon x, 0) }\, \,  e^{i  \epsilon^2 q_1^{(k)} \int_0^x f_1(\epsilon x, 0) dx }\quad \text{ for } k=1,2,3
\end{equation}
where (recalling $\xi = \epsilon x -\epsilon \lambda_1 t$ and $\tau = \epsilon^3 t$)
\begin{align}
\label{eq:f1_example}
    f_1(\xi, \tau)=\frac{12K_{DS}}{K_{NL}} \partial^2_{\xi} \log\bigg[1 + e^{\eta_1} + e^{\eta_2} + \bigg(\frac{k_1 - k_2}{k_1 + k_2}\bigg)^2e^{\eta_1 + \eta_2}\bigg].
\end{align}
The three equations (Eq.~\ref{eq:psi_init}) are evolved according to Eq.~\ref{mnls} by direct numerical simulation. It is interesting to note that the indefinite integral in Eq.~\ref{eq:psi_init} can be explicitly evaluated by the very nature of $f_1(\epsilon x, 0)$ given by Eq.~\ref{eq:f1_example} and is direct consequence of Hirota's bilinear method ~\cite{hirota_2004}. We refer to this solution as \textit{numerical}. On the other hand Eq.~\ref{nsolitonkdvoriginal} in combination with Eq.~\ref{eq:hydro_f} represents an analytical expression for the fields $\psi_j(x,t)$ that are approximate solutions to Eq.~\ref{mnls}. We refer to this expression as \textit{analytical}. \\

In the top panel of Fig.~\ref{fig:intra_chiral} we show a comparison of the density of the first component following the chiral dynamics (Eq.~\ref{derivedKdV}) and the original dynamics (Eq.~\ref{mnls}). 
The time evolution is shown in the \emph{travelling frame of reference} (where the chosen eigenvalue $\lambda_1$ is the speed of the moving frame). In this frame, the two peaks (of different heights and widths) move in opposite directions (as depicted by the arrows). As time progresses, these peaks merge and then separate. In other words, they pass through each other. This is expected of a true KdV multi-soliton however we find that this is almost the case even with the solution of the original non-integrable MNLS (Eq.~\ref{mnls}), \textit{i.e.} we find very good agreement between the two dynamics (see top panel of Fig.~\ref{fig:intra_chiral} which shows the various time snapshots). A three-dimensional plot showing these coherent objects passing through each other is shown in the bottom left of Fig.~\ref{fig:intra_chiral}. To get a better insight into the world-line of the soliton-like objects we also show how the position of the peaks move in time (bottom right of Fig.~\ref{fig:intra_chiral}). Therefore in Fig.~\ref{fig:intra_chiral}, we have successfully shown that we can engineer optimal initial conditions which when evolved according to Eq.~\ref{mnls}, behave almost as solitons although Eq.~\ref{mnls} is not integrable. 


\begin{figure}
\centering
    \includegraphics[width=0.80\textwidth]{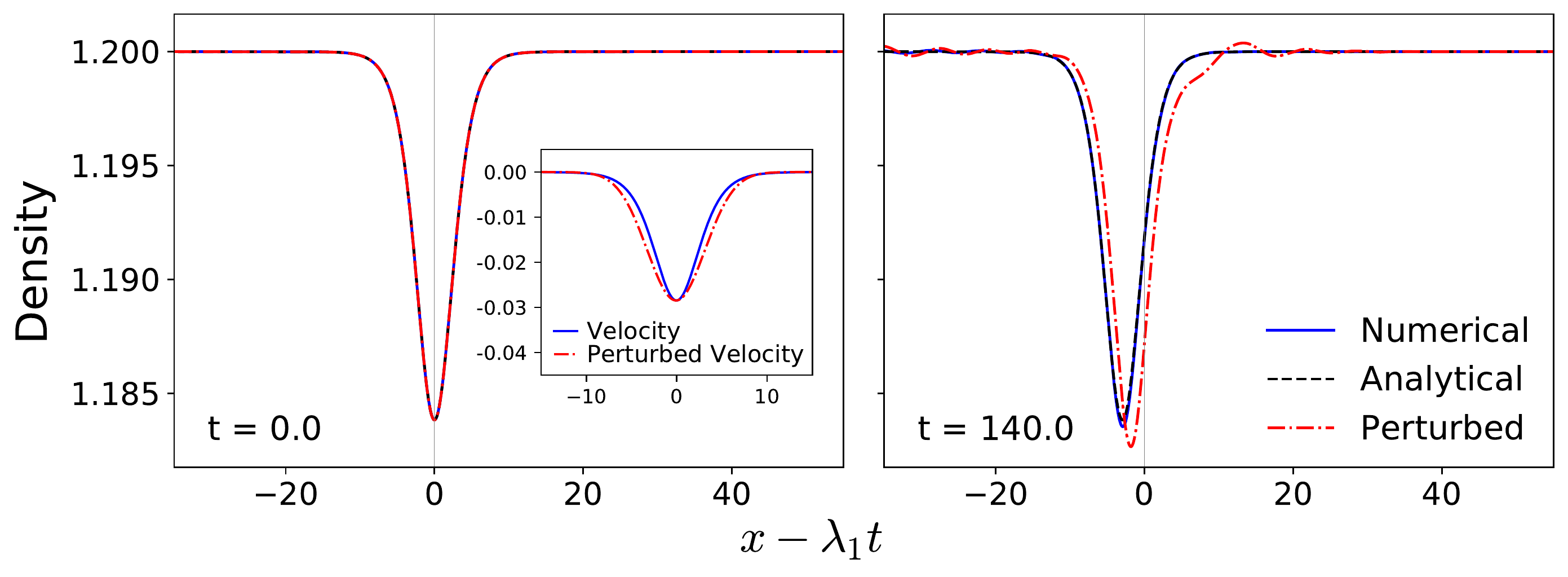}
    \includegraphics[width=0.80\textwidth]{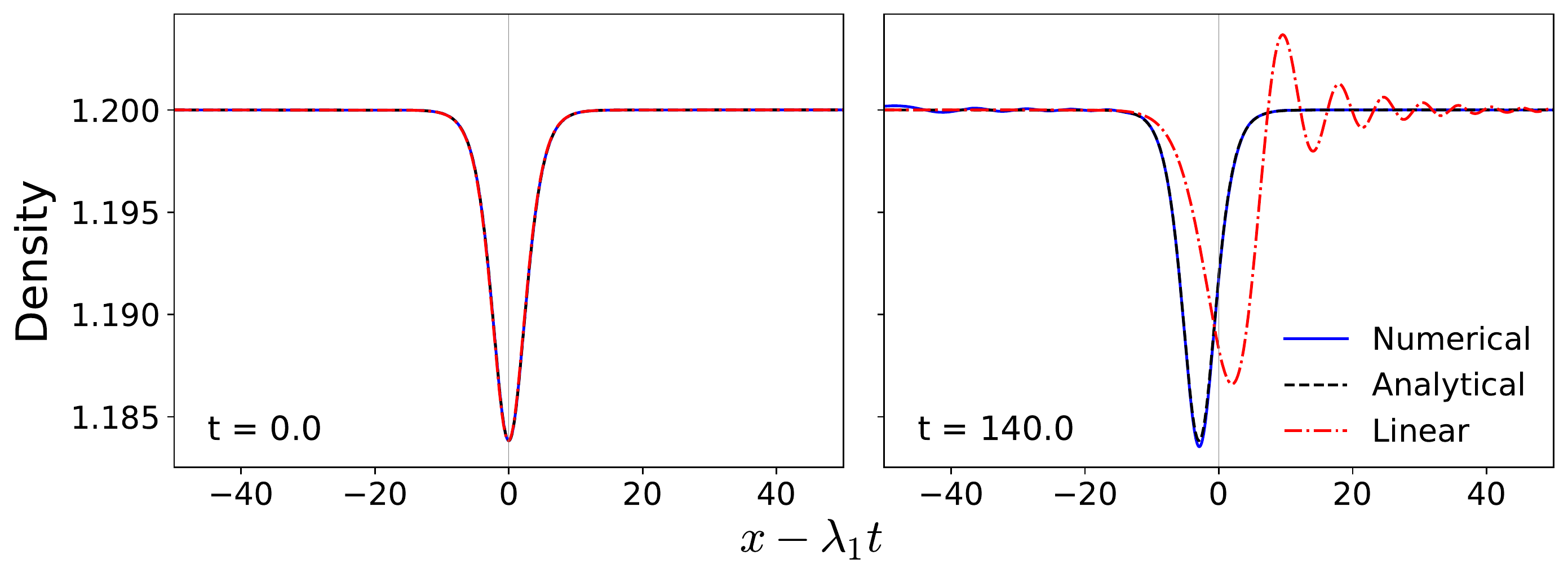}  
    \caption{\footnotesize (colour on-line) (Top) Comparison of the time evolution of the initial profile from the effective description and a slightly perturbed initial profile. 
    The perturbed initial profile is made by using a Gaussian of same height and a slightly wider base for the velocity profile (see inset). The plots above are for two component ($N=2$) system and the velocity components are given by $v_1 = -0.0284 \times \exp{[-x^2/(2\times3.2^2)]}$ and $v_2 = -0.0391 \times \exp{[-x^2/(2\times3.2^2)]}$. The $L-$infinity norm of the difference comes out to be $0.0047$ and $0.0049$ respectively.
    (Bottom) Comparison of the effective chiral description with the linear order system (i.e., time evolution of Eq. \ref{eq:hydro_lin}). 
    For both the plots, we choose the following values for the parameters: $g_1=1, \; g_2=1.2,\;h=0.5,\; \epsilon=0.2,\; \rho_{01}=1.2,\; \rho_{02}=1.4,\; G_1=0.8,\;G_2=0.6,\;dx= 0.046875$ and $dt= 5.0\times10^{-4}$. 
    The soliton parameters are chosen as $k_1=1.4,\;\eta_{1}^{0}=0.0$.} \label{fig:perturbed_linear}
\end{figure}


\subsection{Remarks about ad hoc initial conditions and  nonlinearity}
\label{sec:adhoc_linear}

From the discussions in the previous Sec.~\ref{sec:intra}, two important questions naturally arise. The first question is whether one needs to go through the systematic procedure discussed in the paper to engineer optimal initial conditions. In other words, one might wonder what would happen if we choose initial conditions which seem reasonable based on physical insights. The second question that arises is how nonlinear the dynamics presented in Fig.~\ref{fig:intra_chiral} actually is. Another way of posing this question is whether these coherent excitations are indeed non-trivial or are they merely pulses evolving according to a linearised version of Eq.~\ref{mnls} or equivalently a linearised version of Eq.~\ref{mcqv}. 

To answer the first question, we consider a rather extreme scenario.  We demonstrate that not only ad hoc initial conditions (ad hoc but still based on physical insights and respecting some basic properties) are doomed to fail but also even minor deviations from the optimal initial conditions (derived by our systematic procedure) lead to significant radiative effects. 

As an example, in top panel of Fig.~\ref{fig:perturbed_linear}, we have chosen a two component ($N=2$) MNLS with a one-soliton ($M=1$) initial condition. The time evolution of the density of the first component is shown. The initial density profile is the same as the one obtained by our procedure. However the velocity profile is slightly perturbed. Indeed this velocity perturbation results in the density evolution significantly differing from the  solution for the optimal initial condition, in the sense that, we see far more radiation in the former (see top right panel of Fig.~\ref{fig:perturbed_linear} at a particular late time snapshot). In addition to radiation, the size (depth, width) and location of the peak is also different. This highlights the importance of our systematic analysis in engineering robust propagating initial profiles for the original dynamics (Eq.~\ref{mnls}).



To answer the second question we compare nonlinear (Eq.~\ref{mnls} or equivalently Eq.~\ref{mcqv}) and linear dynamics. For the linear dynamics, we consider the following ($k=1,2,...N$) linearised version of Eq.~\ref{mcqv}
\begin{equation}
\partial_t\delta\rho_k + \rho_{0k}\partial_x\delta v_k = 0,\quad \partial_t \delta v_k + \partial_x\left(\sum_{j=1}^{N}\alpha_{kj}\delta \rho_j + 2G_{k} \rho_{0k}\delta\rho_k\right) +  \frac{1}{4 \rho_{0k}}\frac{\partial^3 \delta\rho_k}{\partial x^2}  = 0.
\label{eq:hydro_lin}
\end{equation}

The bottom panel of Fig.~\ref{fig:perturbed_linear} shows a clear difference between the true nonlinear and the linear dynamics (when we evolve our carefully engineered initial condition in both cases). We see that the linear dynamics predicts the incorrect peak position and is plagued by significant dispersive effects. In fact, in the nonlinear dynamics, it is precisely the intricate balance between nonlinearity and dispersion that results in robust soliton-like behaviour even in a non-integrable model.

\subsection{Bidirectional initial profile: Combination of KdV equations of both sectors}
\label{sec:inter}

%
Till now we only discussed situations when we fix a particular moving frame. In other words, we pick a simple eigenvalue $\lambda_j$ and thereby obtain a single specific KdV equation (Eq.~\ref{derivedKdV}). Said another way, we focus on perturbations propagating with a single sound speed. In this section, we prepare an initial condition that is completely outside the paradigm of our paper and the scheme discussed. Here, we  prepare the initial profile, for our original dynamics (Eq.~\ref{mnls}), derived from the soliton solutions of two \textit{different} chiral equations. Note, since KdV is an equation with only one time derivative, it cannot model the interaction of disturbances propagating in opposite directions. More importantly there is no \textit{a priori} reason to suspect the asymptotic analysis, which gave us approximate solutions and optimal initial conditions, will continue to hold true when initial conditions are prepared using different chiral sectors.

\begin{figure}[h]
    \includegraphics[width=\textwidth]{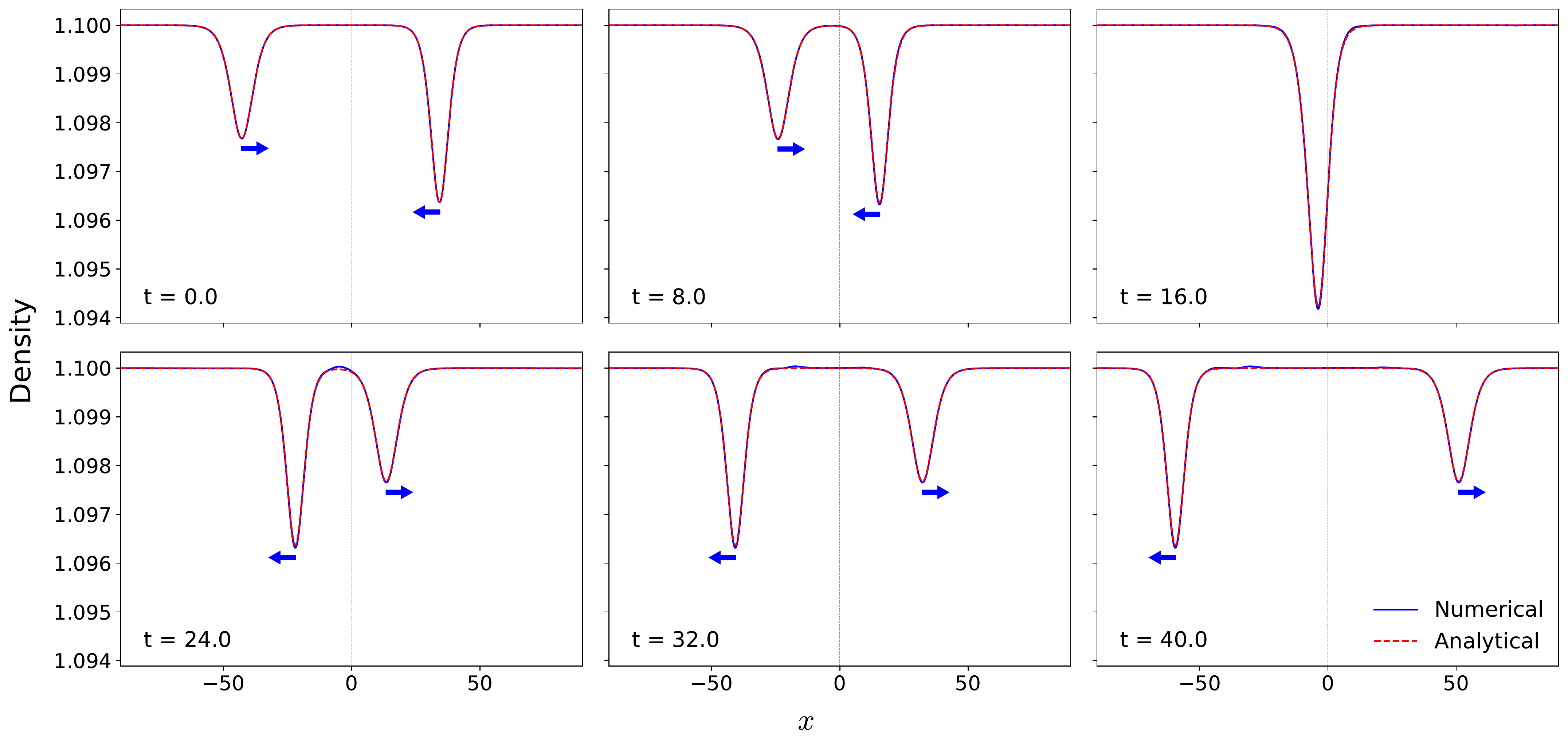} 
    \vspace{4mm}
    \includegraphics[width=.54\textwidth]{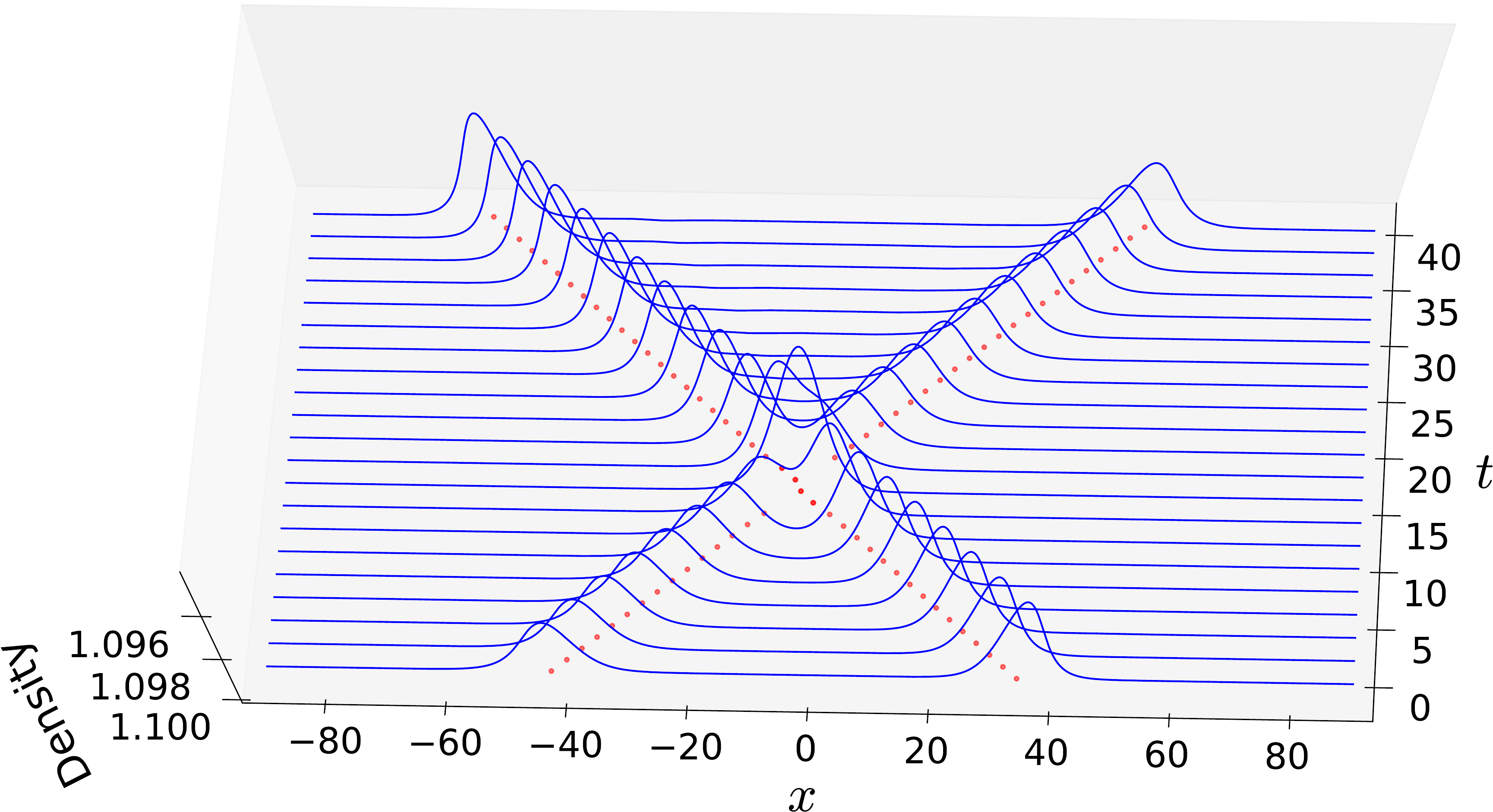} 
    \hspace{1mm}
    \includegraphics[width=.44\textwidth]{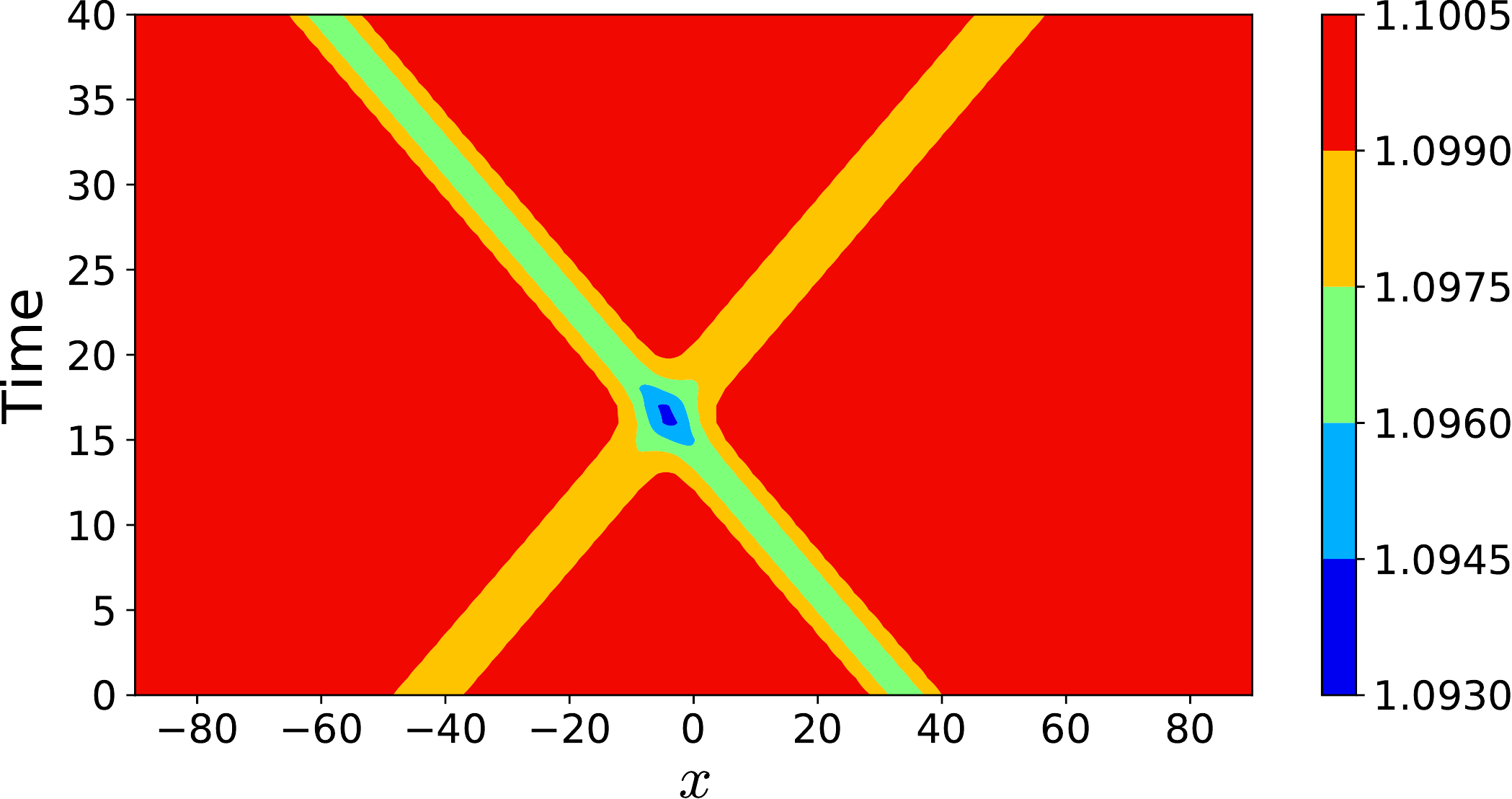}
    \caption{\footnotesize (colour on-line) Time evolution of an artificial initial condition created by joining two opposite travelling well separated peaks (Eq.~\ref{eq:hydro_f_inter}). 
    (Top) Comparison of the numerical evolution of density of the original dynamics (Eq. \ref{mnls}) with the artificial initial condition and the linear superposition of two oppositely travelling KdV solitons (i.e., solitons from $j=1$ and $j=6$ in Eq.~\ref{derivedKdV}). 
    The arrows indicates the direction of the two soliton peaks. 
    (Bottom) Time evolution of the two soliton peaks with the original dynamics. Bottom left shows a three dimensional plot (along with the peak positions marked in red) and bottom right shows the contour plot of the same figure. 
    For the plots, we chose the following values for the parameters: $g_1=1, \; g_2=1.2,\; g_3=1.3,\;h=0.5,\; \epsilon=0.2,\; \rho_{01}=1.1,\; \rho_{02}=1.2,\;\rho_{03}=1.4,\; G_1=0.7,\;G_2=0.9,\;G_3=0.8$,  $dx= 0.046875$ and $dt= 5.0\times10^{-4}$. 
    The soliton parameters are chosen as $k_1=0.8,\;\eta_{1}^{0}=-15.0$ for soliton corresponding to $\lambda_1$ (left moving) and $k_1=1.0,\;\eta_{1}^{0}=15.0$ for soliton corresponding to $\lambda_6$ (right moving).
    The plots show time evolution of the first component of the three component system ($N=3$) evolved using the finite difference scheme given by Eq.~\ref{discretizedmnls}.}
    \label{fig:inter_chiral}
\end{figure}

Let us consider the three component ($N=3$) MNLS with a one-soliton ($M=1$). The opposite speeds (simple eigenvalues) are $\lambda_1$ (right moving) and $\lambda_6$ (left moving). This means that we have two single-soliton solutions (see Eq.~\ref{eq:msol})  $f_1(\xi,t)$ and $f_6(\xi,t)$. We combine these two solutions to prepare an initial condition for the original MNLS (Eq.~\ref{mnls})

\begin{align}
		\begin{pmatrix}
		\vec{\rho}(x,0)\\
		\vec{v}(x,0)
		\end{pmatrix} 
		= 
		\begin{pmatrix}
		\vec{\rho}_0 \\
		0
		\end{pmatrix}
		+ \epsilon^2  f_1(\xi, \tau)V_1 +  \epsilon^2  f_6(\xi, \tau)V_6\quad \text{[Bidirectional Initial Condition]}
		\label{eq:hydro_f_inter}
	\end{align}
where $V_1$ and $V_6$ in Eq.~\ref{eq:hydro_f_inter} can be obtained using Eq.~\ref{eq:hydro_f}. We re-emphasize that preparing a bidirectional initial profile as shown in Eq.~\ref{eq:hydro_f_inter} does not fall under the paradigm of our effective chiral reduction. Nevertheless we do this to test the limits of our constructions. To our surprise the two opposite moving excitations behave as if they are solitons (see Fig.~\ref{fig:inter_chiral}) passing through one another with minimal radiative loss. This is a completely unexpected phenomenon to which we afford no theoretical explanation presently but only record that it seems a generic feature of our prescription for MNLS dynamics, unaffected by the choice of coupling coefficients and background densities. 


%
%

\section{Repeated sound speeds}
\label{sec:CRSS}

Thus far in this paper we limited the discussion to the case when all eigenvalues of $\mathcal A$ are simple.  As a result we obtained a single KdV equation (Eq.~\ref{derivedKdV}) once a specific eigenvalue $\lambda_j$ was chosen. This eigenvalue (and the coupling coefficients and background densities) allowed us to construct the optimal initial conditions for robust propagation. We then compared the resulting approximate time dynamics with that dictated by original equation (Eq.~\ref{mnls}). The alternative and, in our opinion, fascinating phenomenon where $\mathcal A$ (Eq.~\ref{eq:matA}) admits repeated eigenvalues will be part of a subsequent work. Here we simply note the resulting chiral equations take the form of a system of coupled KdV-type equations and present a non-trivial generalisation of the results in the present manuscript. In this section, we will lay down the foundation for the subsequent investigation. We first put forward the following important theorem. 

{\theorem{Suppose $\lambda^2$ is an eigenvalue of $\rho \tilde{\alpha}$. For $m\geq 2$, the multiplicity of $\lambda^2$ is $m$ if and only if the eigenvalue takes the form $\lambda^2 = \rho_{0i}(\tilde{g}_i-h)$ for $m+1$ pairs $(\rho_{0i}, \tilde{g}_i)$ .}}\\

We do not present the proof of this theorem here since it is not particularly illuminating. We remark in passing that this peculiar property of $\rho\tilde\alpha$ is entirely a consequence of the specific form for the cross-species coupling assumed. The main import of the theorem is that any time a repeated eigenvalue occurs, the eigenvalue must be of the form specified. This then implies all coefficients of the associated reduced KdV equations are fully defined only in terms of the coupling coefficients and background densities.

 Indeed there are additional consequences of the aforementioned condition on $\lambda^2$. Since there are $m+1$ $(\rho_{0i},\tilde g_i)$ pairs involved in the theorem, without loss of generality, let us number them $i=1,2,\ldots,m+1$. This is permissible since each species is coupled in exactly the same manner to every other species. Only the self-interaction $\tilde g_i$ distinguishes species. Hence we have
\begin{align}
\lambda^2 = \rho_{01}(\tilde{g}_1 - h) = \ldots = \rho_{0m+1}(\tilde{g}_{m+1}-h).\label{eqn:equal_evals}
\end{align}
Note that the above expression is true for a particular $h$. Thus eliminating $h$ from any two $(\rho_{0i},\tilde g_i)$ pairs we obtain multiple expressions for this value 
\begin{align}
h = h_{ij} = \frac{\rho_{0i} \tilde{g}_i-\rho_{0j} \tilde{g}_j}{\rho_{0i}-\rho_{0j}},\quad i\neq j,\,\:i,j\in\{1,2,\ldots,m+1\}.\label{eqn:h_condition}
\end{align}
\noindent By definition $h_{ij}=h_{ji}$. Next observe that by equating any two $h_{ij}$ we obtain a constraint between various $\tilde{g}_i,\rho_{0i}$. One might naively expect a large number of such constraints however many of these are redundant in the sense that
\begin{align}
\label{eq:hijhik}
h_{ij}=h_{jk} \iff h_{ik} = h_{kj} \iff h_{ij} = h_{ik}.
\end{align}
Indeed all three of the above equalities (Eq.~\ref{eq:hijhik}) reduce to precisely one constraint
\begin{align}
    \rho_{0i} \tilde{g}_i (\rho_{0j}-\rho_{0k}) + \rho_{0j} \tilde{g}_j(\rho_{0k}-\rho_{0i}) + \rho_{0k} \tilde{g}_k (\rho_{0i} - \rho_{0j}) &=0.
\end{align}
Thus a minimal set of equalities between $h_{ij}$ which imply all others in Eq.~\ref{eqn:h_condition} is 
\begin{equation}
\label{eq:h12h23}
h_{12}=h_{23},\quad h_{23} = h_{34},\quad\ldots\quad h_{m-1\: m}=h_{m\: m+1}.
\end{equation}
The above Eq.~\ref{eq:h12h23} is equivalent to the following $(m-1)$ constraint equations
\begin{align}
\label{eq:r1f}
    \rho_{01} \tilde{g}_1 (\rho_{02}-\rho_{03}) + \rho_{02} \tilde{g}_2(\rho_{03}-\rho_{01}) + \rho_{03} \tilde{g}_{3} (\rho_{01} - \rho_{02}) &=0\\
    \rho_{02} \tilde{g}_2 (\rho_{03}-\rho_{04}) + \rho_{03} \tilde{g}_3(\rho_{04}-\rho_{02}) + \rho_{04} \tilde{g}_4 (\rho_{02} - \rho_{03}) &=0\\ \nonumber
    &\vdots\\
    \rho_{0m-1} \tilde{g}_{m-1} (\rho_{0m}-\rho_{0m+1}) + \rho_{0m} \tilde{g}_{m}(\rho_{0m+1}-\rho_{0m-1}) + \rho_{0m+1} \tilde{g}_{m+1} (\rho_{0{m-1}} - \rho_{0m}) &=0.
\label{eq:rmlast}
\end{align}
\noindent The above system of equations represents a necessary condition for repeated eigenvalues (repeated sound speeds) to exist in our system. This condition is given in terms of the coupling coefficients and background densities. Irrespective of the size of the system $N$, if a repeated eigenvalue exists then some subset of $m+1$ pairs $(\rho_{0i},\tilde g_i)$ satisfy the above $(m-1)$ equations. Thus if no subset of the $(\rho_{0i},\tilde g_i)$ satisfy the above relations, then the system consists of only simple eigenvalues and the results from the earlier part of the current paper are operative. 

Eqns.~\ref{eq:r1f} - \ref{eq:rmlast} are also a sufficient condition for repeated eigenvalues. Indeed if one selects the coupling constants and background densities to satisfy the above constraint, then a little algebra shows one can define $h$ such that Eq.~\ref{eqn:h_condition} holds. If this $h$ is positive, then one has indeed found a value of the cross-coupling constant to guarantee a repeated eigenvalue of $\rho \tilde{\alpha}$. Furthermore, from Eq.~\ref{eqn:equal_evals}, this repeated eigenvalue (of multiplicity $m$) is given by $\lambda^2 = \rho_{0i}(\tilde{g}_i-h)$ where one may use any of the $i=1,\ldots,m+1$. 

Since there are $(m-1)$ equations in $2(m+1)$ variables, one can always find a solution. Evidently there is a $2(m+1)-(m-1)=m+3$ dimensional real manifold such that each point on the manifold is a potential system to guarantee repeated eigenvalues. Thus there are in fact many ways to construct a system with repeated eigenvalues. Solving Eqns.~\ref{eq:r1f} - \ref{eq:rmlast} is also straightforward. For instance one can pick any $\rho_{0i}>0$ and then consider the constraints as a linear system of equations for the $\tilde g_i$. The vector of $\tilde g_i,\: i=1,2,\ldots m+1$ belong to the null space of matrix. One can show this matrix generically has a two-dimensional null-space (when all $\rho_{0i}$ are unequal).

For each such choice of $(\rho_{0i},\tilde g_i)$, there is a unique $h$ that gives rise to the repeated eigenvalue. If $h$ is varied even slightly from this critical value (while keeping $(\rho_{0i},\tilde g_i)$ fixed) then the repeated eigenvalue splits up into unequal eigenvalues. Nevertheless the prescription described in this section gives the experimentalist the precise value of the cross-component coupling $h$ and that will guarantee repeated sound speeds in a very transparent manner. This also closes a gap in the analysis of our previous work \cite{Swarup2020} where we stated multiple eigenvalues were possible for specific values of $h$ but were not able to provide a full description of that scenario.


\section{Conclusions and Outlook}

In this paper, we addressed a general question of the possibility of constructing initial conditions for generic non-integrable models, such that they bear as much resemblance as possible to solitons of integrable models. We successfully found such localised excitations that move at almost constant speed and barely show scattering / radiation effects. Our construction is systematic and its success was demonstrated in Fig.~\ref{fig:intra_chiral}. We also discussed two natural questions -  (i) Can one make crude attempts to design initial conditions by circumventing our systematic prescription?  and (ii) How truly nonlinear is our dynamics? We provided convincing evidence that (i) Crude attempts to design initial conditions by evading our procedure is doomed to fail  (top panel of Fig.~\ref{fig:perturbed_linear}) and (ii) Our designed initial conditions undergo truly nonlinear evolution (bottom panel of Fig.~\ref{fig:perturbed_linear}). We also presented an interesting finding on bidirectional evolution that is composed of both chiral sectors (Fig.~\ref{fig:inter_chiral}). As this falls outside the paradigm of our formalism we present this as an interesting observation but no theoretical explanation is offered. It is indeed remarkable that one can find excitations moving in opposite directions (for a non-integrable model such as Eq.~\ref{mnls}) that have strong resemblance with solitons of integrable models. Needless to mention, the findings in this paper can serve as a guiding principle to engineer initial conditions (localised excitations) in experiments (such as cold atoms or nonlinear optics) that can subsequently display robust soliton-like evolution. Indeed our expressions may also serve as suitable initial guesses for numerical routines employing the Ansatz-based approach to obtain travelling-wave excitations. The expressions and the resulting dynamics also strongly suggest there do in fact exist stably propagating localised solutions to systems of coupled PDEs such as Eq.~\ref{mnls}.

In the future, we plan to investigate the case of degeneracy (repeated eigenvalues), the foundation for which has already been laid out in this paper (Sec.~\ref{sec:CRSS}). This is naturally expected to result in coupled KdV equations which might in turn give rise to possibility of finding new integrable chiral field-theories (non-trivial generalisations to Eq.~\ref{derivedKdV}). It is also paramount to mention that although, we used MNLS (Eq.~\ref{mnls}) as a platform, our formalism can be exploited to understand nonlinear dynamics in several other models, such as classical spin chains~\cite{lakshmanan2011fascinating, lakshmanan1976dynamics, tjon1977solitons, das2020nonlinear,porsezian1992integrability} which too is a subject of future investigation.

\label{sec:C}

\section*{Acknowledgements}

We thank Ziad Musslimani, Andrea Trombettoni, Chiara D'Errico, Nicolas Pavloff, Bernard Deconinck, Konstantinos Makris and Swetlana Swarup for useful discussions. MK would like to acknowledge support from the project 6004-1 of the Indo-French Centre for the Promotion of Advanced Research (IFCPAR), Ramanujan Fellowship (SB/S2/RJN-114/2016), SERB Early Career Research Award (ECR/2018/002085) and SERB Matrics Grant (MTR/2019/001101) from the Science and Engineering Research Board (SERB), Department of Science and Technology, Government of India. VV would like to acknowledge support from SERB Matrics Grant (MTR/2019/000609) from the Science and Engineering Research Board (SERB), Department of Science and Technology, Government of India.

\section*{References}
\bibliography{Bibliography}
\end{document}